# MoS2 Based 2D Material Photodetector Array with high Pixel Density

Russell L. T. Schwartz[1,2], Hao Wang[1,2], Chandraman Patil[1,2], Martin Thomaschewski[3], Volker J. Sorger[1,2]*

[1]Department of Electrical and Computer Engineering, University of Florida, Gainesville, FL, 32611, USA

[2]Florida Semiconductor Institute, University of Florida, Gainesville, FL, 32611, USA

[3] Thomas J. Watson Laboratory of Applied Physics, California Institute of Technology, Pasadena, CA, USA

**Abstract: Arrays of photodetector-based pixel sensors are ubiquitous in modern devices, such as smart phone cameras, automobiles, drones, laptops etc. Two dimensional (2D) material-based photodetector arrays are a relevant candidate, especially for applications demanding planar formfactors. However, shortcomings in pixel density and prototyping without cross contamination limit technology adoption and impact. Also, while 2D material detectors offer high absorption, graphene's closed bandgap results in undesirably high dark currents. Here, we introduce the experimental demonstration of dense planar photodetector arrays. We demonstrate a micrometer narrow pitched 2D detector pixels and show this approach's repeatability by verifying performance of a 16-pixel array. Such dense and repeatable detector realization is enabled by a novel, selective, contamination free 2D material transfer system, that we report here in automated operation. The so realized photodetectors responsivity peaks at 0.8 A/W. Furthermore, we achieve uniform detector performance via bias voltage tuning calibration to maximize deployment. Finally, we demonstrate 2D arrayed photodetectors not only on a silicon chip platform but also demonstrate and very array performance on flexible polymer substrates. Densely arrayed, flat, bendable, and uniform performing photodetector pixels enable emerging technologies in the space where lightweight and reliable performance is required, such as for smart phones and emerging AR/VR markets, but smart gadgets, wearables, and for SWAP constrained aviation and space platforms.**

# Main Text:

## 1. Introduction

Image sensors have abundant applications in today's world from biomedical applications [1-2] to smart devices internet of things (IoT) [3], focal plane arrays [4], to wearable devices [5]. Especially, for network edge applications a reduced formfactor is of relevance. Thin film image sensors offer an added advantage in these fields by having a low impact on space constraints and lowered power consumption, making them easier to integrate [6-7]. Two-dimensional (2D) materials have attracted attention due to their unique properties physically, electrically, and optically [8-11]. As a class of materials transition metal dichalcogenides (TMDCs) have been extensively researched for their usage in photodetector devices [12-15]. Molybdenum disulfide ($MoS_2$) has emerged as a promising candidate due to its direct bandgap in monolayer form, layer dependent electrical and optical characteristics, high absorption in the visible range, and ease of integration with different material systems [16-30]. A variety of devices and systems have been shown with MoS$_2$ including photoconductive detectors, photo-gated detectors, and detector arrays/Image sensors. However, these systems still have issues generally with large pitch spacings between detectors, non-uniform device performance, high-voltage requirements, and fabrication incompatibilities with desirable materials.

Here, we present an MoS$_2$ based photodetector array with a dense sub-10-micron pitch in both lateral directions with a maximum responsivity of 0.8 A/W that can be fabricated on nearly any substrate. The arrays are shown on a standard rigid substrate and on a flexible polymer substrate with operation continuing to a maximum bending radius of 1.25 cm and less than 10% variation of device performance across repeated bending cycles. Also demonstrated is a post-fabrication tuning process to ensure that all devices in the system operate at the same responsivity level using a maximum of 2 V bias. This is achieved by careful design of the array electrodes and the usage of a specialized 2D material transfer printer [31-32] to accurately transfer exfoliated flakes of MoS$_2$ to designated areas on the electrode design. With the assistance of a semi-automated program an operator of the system can transfer flakes on to the design with an average rate of 1 per minute. We believe that these arrays can see potential usage in wearable devices by being resilient to bending conditions, having a high density, and reliable performance with tuning.

## 2. Array Design

The arrays were designed with MoS$_2$ as the active photo-sensitive material, with a peak response at 670 nm, the arrays are intended to operate in visible light regime, mainly in the red portion of the spectrum. Each detector in the array is a two-electrode device operating as a photoconductive detector where the incident light generates electron-hole pairs (EHPs) contributing to the photocurrent of the device (Fig 1-a.). Photoconductive detectors were chosen over phototransistors to maximize the density of the devices rather than a focus on device performance, allowing the array to fabricated in only two layers by a process conducive to rapid prototyping and allowing an avenue for high-throughput manufacturing. To maximize density of the arrays they were designed such that the device pitch in both horizontal and vertical directions were equal creating a densely packed square array of devices. The pitches tested for these arrays scale from 50 μm to 9.5 μm, with each pitch corresponding to a specific spacing between the electrodes denoted as the device width (Fig 1-bc.). In addition to the pitch the volumetric density is ruled by the maximum height of the array above the substrate, this can be found by a simple addition of the electrode height and the height of the thickest MoS$_2$ flakes in each array. With MoS$_2$ as a 2D material device thickness remained low ranging from below 100 nm to a maximum of 160 nm including the 50 nm electrode lines. The arrays then have a maximum pixel density of 400,00 pixels/cm$^{-2}$ with total thickness below 160 nm creating a densely packed array with respect to volume. Electrodes were designed to be prefabricated with MoS$_2$ flake placement occurring later, with all devices sharing common ground lines and each with a separate signal/supply line allowing for individual post-tuning with control over the bias voltage. The design of the arrays is intentionally substrate independent such that arrays can be fabricated to conform to the material system of their eventual application. Notably the arrays were fabricated on SiO$_2$ substrates and a flexible polymer substrate, polyimide (PI), giving rise to potential applications in the wearable devices field.

## 3. Results & Discussion

Devices are first characterized for their performance as photoconductive detectors and then characterized as whole array for variations across devices, tuning characteristics, and post-tuning variations. Focus is placed on the characterization for operational conditions of the device for a flexible device and preparation conditions of the array impacting the device performance.

### 3.1 Device Characterization

The devices are tested over the visible spectrum from 500-900 nm at varying optical power levels to see the photo response with the corresponding conditions. Results for these tests show a main peak at 670 nm which increases in relative strength with decreasing optical power level, up to a maximum responsivity of 0.8 A/W (Fig. 2-a). This suggests that the device is saturating with higher optical power levels, further evidenced by the response curves shown on average decreasing in responsivity with the increasing optical power levels. The primary peak seen in the response corresponds well with main excitonic peak in $MoS_2$ which has been shown to exist in the region of 1.8-1.9 eV. In addition to the main peak seen at 670 nm a secondary peak can be seen around 600 nm corresponding to the secondary excitonic peak around 2.0 eV. Finally, a third peak at 860 nm is caused by indirect band gap absorption seen in multilayered materials. Another pertinent result from this testing is the decrease seen in the main peak with increasing optical power, suggesting that the saturation of the devices is significant in the direct band transition, however the secondary excitonic peak remains relatively constant throughout shifting optical power levels. Additionally, the indirect band transition sees a significant initial decrease from increasing optical power but less of a decrease at further increasing power levels (Fig. 2-b). This decrease however does not affect the total function of the device, with the main response shifting to slightly shorter wavelengths in visible spectrum with increased optical power. Furthermore, the red-light regime can still be detected albeit with a lower response than the red-yellow border regime, and the near infrared around the indirect peak remains relatively stable.

Exploring further the device characteristics with changing optical power we observe that as the optical power is even further increased, to microwatt levels, that a roll off is occurring in the device saturation in the main peak (Fig. 2-b.). Additionally, it is shown that the device saturation characteristics are occurring at an earlier power level in the indirect gap transition, where roll off is beginning to be seen in the 10's-100's of nanowatt range (Fig. 2-b.). Finally, in the second excitonic peak a relatively consistent response is seen suggesting there's no significant saturation occurring, although it is possible that the saturation still occurs outside of the tested optical power ranges. This is supported by the fact that the optical power range at the 600 nm wavelength test is slightly lower than that of 670 and 680 nm. Further the upper limit of the 600 nm power range appears to begin dropping in responsivity but needs a larger test span for conclusive evidence.

Testing is also done on device characteristics for applications in potential applications in flexible substrates and systems, such as smart clothing, for example. The testing for this is executed using a polyimide substrate which is subsequently exposed to various bending conditions. The initial bending tests are done to find the maximum bending radius under which the devices would cease to operate properly, defined as follows; this is done by testing for device characteristics of a freshly fabricated array and then subjecting the array to various levels of bending and retesting to see how the characteristics of the device changed. Results for this line of testing yielded a maximum bending radius around 1.25 cm, with a slight increase in the response of the devices with a small bending radius 4.5-5.0 cm and a decrease towards the maximum bending radius (fig. 2-c.). The curve for this type of response with respect to bending radius fits closely with a negative second order polynomial function with the intercept set at the unbent level, with an $R^2$ value of 0.997. One explanation for the initial increase in responsivity under these test conditions is that the uniaxial strain is shifting the peak of the band gap and thus the optical response to match the testing wavelength of 680 nm. The $MoS_2$ band gap is known to have a strong dependency on strain with a 45-70 meV redshift happening per degree of strain [12, 33-34]. To shift the main excitonic peak it would be 27 meV or roughly 0.5% strain, and to shift the secondary excitonic peak it would be a 216 meV or roughly 3-4% strain. With a 3 mm-thick substrate and a 4.5 cm bending radius the strain induced on the $MoS_2$ is 3.3% making it possible for the increase to be from the secondary excitonic peak shifted to be at the 680 nm testing wavelength. Furthermore, the reduced responsivity at mA/W levels is because testing was done with a laser operating around 10 µW approximately 2 orders of magnitude higher than the start of saturation that was seen previously. With the maximum bending radius established it is then used for stress testing over cycles between flat and the maximum bending radius, where the device characteristics are tracked over 100 cycles. The arrays experience relatively little change when tracked over the 100 cycles, having less than a 10% change from the average responsivity after the initial bending. The results for responsivity under bending conditions are promising for usage in flexible systems, with special care being given to not exceed the maximum bending radius of the array.

Devices are also characterized under various preparation conditions, mainly by changing the annealing temperature and time for the devices. Both $SiO_2$ and PI substrates were tested, PI has a glass transition temperature around 250°C leading to a maximum annealing temperature of 200°C to avoid deformation. Previous work has shown that thermal annealing does not help the interface between $MoS_2$ and gold contacts [35], thus the main changes that can be seen in device performance is due to annealing affecting the $MoS_2$ directly. Thermal $MoS_2$ annealing has been shown to have 2 distinct regions where device improvements are shown up to 300°C and then subsequent device degradation beyond 300°C towards 400°C in both photodetectors and FETs [36-38]. Here, the devices see a 5x and 3x improvement on $SiO_2$ and PI substrates, respectively, the improvements are consistent with about half of the improvement seen at 100°C anneal compared to the 200°C anneal. These results suggest that annealing can help to improve the devices significantly and thus limiting the options for flexible substrates to those with high glass transition temperatures. Such substrates as PI and thin glass substrates with some potential on polyethylene naphthalate, although it has a glass transition temperature under 200°C.

### 3.2 Photodetector Array Characterization & Tuning

Photodetector array characterization is done by collecting all the device characteristics for a given 4x4 array and then comparing them with one another. I-V curves are collected per device from a range of -2 to 2 V of photoconductor bias in both dark and illuminated conditions (Fig. 3-a.). This data and physical parameters are then used to find a normalized responsivity mapping of devices (Fig. 3-b.). The mapping between devices shows significant variations with responses at 2V and 680 nm ranging from 1 mA/W to 15.5 mA/W. Responsivities can roughly be clustered into three groups low performance (1 to 3.5 mA/W), middle performance (4 to 8 mA/W), and high performance (9 to 16 mA/W), while we again note, that experiments were carried out at laser powers where saturation occurred, else the responsivities are in the 0.x A/W range, see Figure 2b. No significant differences between these three groups can be established due to thickness, size, or position, leading to a conclusion that innate resistive differences in the flakes and potentially contact resistance is leading to responsivity variations. Post-fabrication tuning for the array to generate a uniform photoabsorption condition and reduce the back-end burden on the digital signal processing (DSP) electronics, is done by changing the voltage bias on individual devices. The tuning level is set to be the lowest maximum responsivity of all devices, in this case devices 15 & 16 have 1.0 mA/W responses so that is set as the tuning level. The initial voltage for tuning was set at 1 V as the middle value for operational range of devices, from there the voltage is adjusted in 100 mV increments to reach the tuned responsivity level. Results for this methodology show bias voltages span the full range of 0.1 V to 2 V, with the final voltage tunings resulting in -0.9 V to +1.0 V (Fig. 3-c.). Reduction in voltage in voltage does not present a significant issue and can lead to reduced power consumption for the sensor, but voltage increases would require a method of amplification in a final system or potentially a local annealing method to further improve device performance. However, the maximally tuned up voltages still fall below the standard CMOS image sensor 3.3V level, potentially leading power savings overall using the $MoS_2$ detector array in an image sensor system. Final post-tuning response levels range from 0.9 mA/W to 1.5 mA/W showing much greater uniformity across the devices (Fig. 3-d.). One potential avenue to further increase device uniformity is to use CVD $MoS_2$ rather than exfoliated material, which has been shown to have more uniform material characteristics and ultimately device performance [39]. The challenge with CVD growth is to find a suitable technique that would be amiable with growth on flexible substrates. Currently thermal CVD growth shows minimum temperatures around 500°C, much higher than most flexible substrates can sustain [40-42]. However, recent advancements have shown potential in ALD and PECVD with lower growth temperatures (150-500°C) for TMDCs creating potential for growth on some flexible substrates [43-45].

## 4. Methods & Manufacturing

Array fabrication is done in two stages, the first was to pre-fabricate the electrode array on the given substrate and the second was to transfer the $MoS_2$ flakes and anneal them. The electrode fabrication followed a standard procedure using electron beam lithography to pattern the design, then deposit the electrodes using electron beam physical vapor deposition and lift off to expose the designed electrodes. The second stage is to perform the $MoS_2$ flake transfer. This is done with a semi-automated version of a stamper based 2D material transfer system [31-32] (Fig. 4). The transfer process is semi-automated where the operator loads both the chip and a sheet of PDMS which has mechanically exfoliated flakes on it into the system. From there the system can begin a startup routine where the operator can choose to manually search for flakes on the PDMS or run an automated routine to compile a flake list which uses a camera and an edge detection algorithm to locate flake centroids. With a list of flakes the operator can then choose which flake/s they wish to transfer to the current device/s, the system then begins the near contact routine which does a rough alignment and brings the flake within 25 μm of the chip. Once the near contact is complete the operator can then choose whether to do a fine alignment for tighter pitched devices or directly stamp the flake on to the device, once the transfer is completed the system will move on to the next device following the same process. After all the devices have been transferred the system will stop and follow a rest routine to return to the load/unload stage where the chip and PDMS film can be removed. Following the automated routine with no fine alignment and several devices to be transferred the system has a throughput of about one transfer every minute. The development of this process is focused on fabricating arrays of devices for various pitch sizes, with fine alignment being necessary for lower pitch dimensions.

## 5. Conclusions

Here we have presented an automated 2D manufacturing system that has allowed the dense fabrication of arrays and automation system to achieve a high throughput of devices. The manufacturing process enables fabrication of many device concepts, while maintaining high rate of throughput and the ability to assess device concepts through rapid prototyping. Additionally, we have demonstrated an $MoS_2$ based photodetector array that has been fabricated both on $SiO_2$ and PI substrates with a maximum responsivity of 0.8 A/W and maximum bending radius of 1.25 cm, while demonstrating to our knowledge the densest 2D photodetector array to date, achieving over 400,000 pixels per $cm^2$ with a maximum device thickness of 160 nm. The density of our arrays compares competitively with industry CMOS sensors which have pitches span the range in the single digit micron scale. Further we have demonstrated an ability to use post fabrication bias voltage tuning to achieve high uniformity among devices made with exfoliated flakes We believe that following this methodology that 2D arrays could see usage in the wearable device field, specifically in smart glasses or lenses technology.

# Figures:

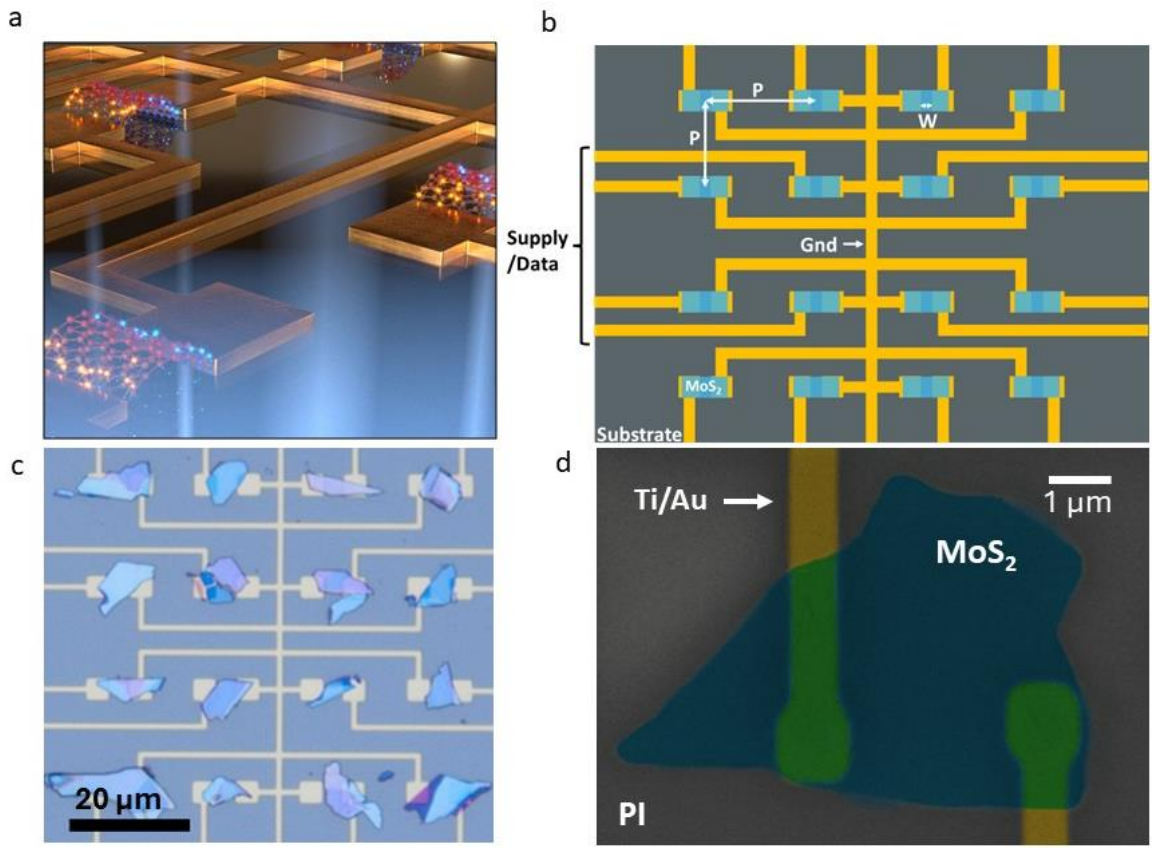

**Fig. 1:** Physical 2D Material (2DM) Array Design. **a**, Conceptual image of 2D array in usage, showing light absorption and EHP generation in the MoS2 devices. **b**, Schematic design of the 2DM array, where p is the pitch of the array in both vertical and horizontal directions, and w is the device width which is dependent on p of the designed array. The design contains a common ground line for all devices in the array and individual signal/source pads for device biasing and current readout. Substrate material is either SiO2 or Polyimide for rigid and flexible applications, respectively. **c**, A fabricated array on $SiO_2$ with pixel pitch of 25 μm, all devices in the array are operational but vary in size and shape from the 2DM selection process. 20 μm scale bar. **d**, False colored SEM image of a device from 9.5 μm pitched array on PI. Titanium (5 nm) and Gold (45 nm) metal lines are shown in gold, while the MoS2 flake for the device is shown in blue, and substrate PI is left uncolored. Scale bar is 1 μm.

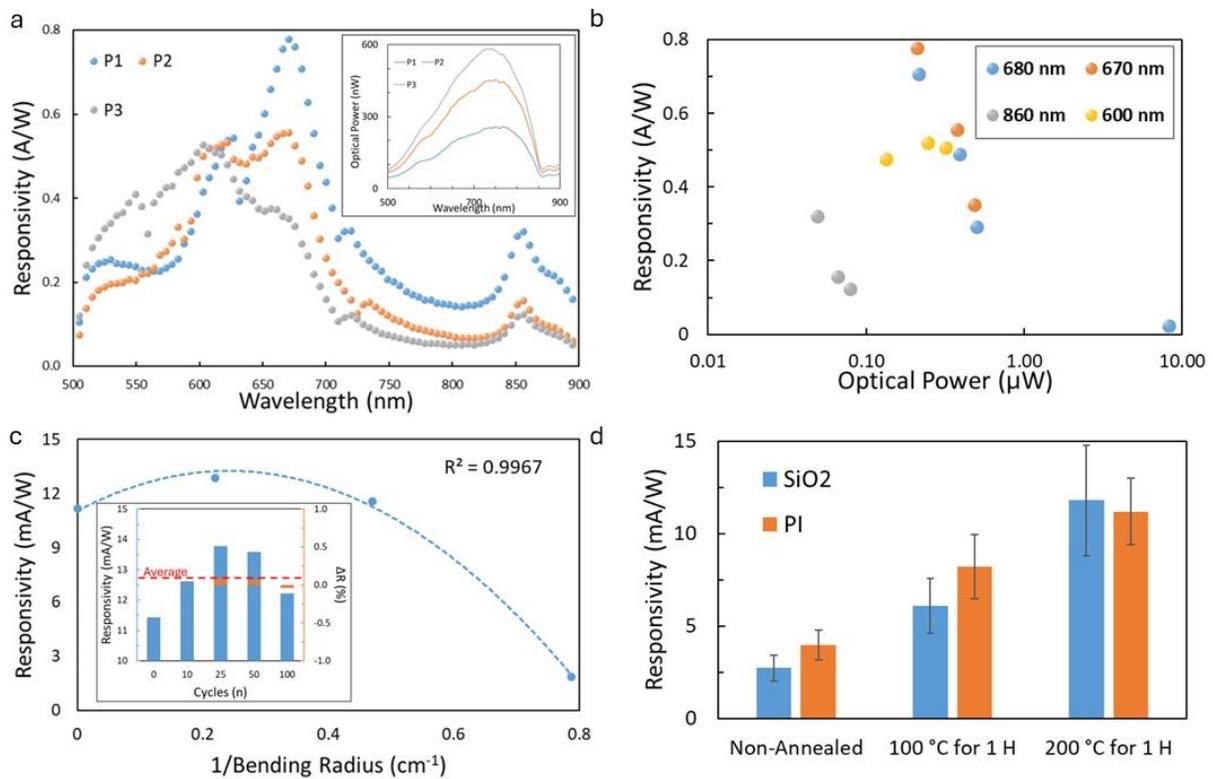

**Fig. 2:** Device and Application Performance. **a**, Responsivity for the optical excitation range of 500 nm to 900 nm, for three different laser power levels. Peak photo response is seen growing at the main peak, 670 nm, with decreasing optical power. Further two additional peaks can be seen 610 nm and 860 nm, corresponding to additional photo response peaks seen in multi-layered MoS2. Inset shows the incident optical power for all three power levels, ranging from 50 nW to 580 nW. **b**, Responsivity over changing optical power at 600, 670, 680, and 860 nm. **c**, Device responsivity over bending radius, slight increases can be seen with shallow bending radii before decreasing to a maximum bending radius of 1.25 cm. The maximum bending radius was used for cyclability testing between flat and bent positions. Inset shows results over 100 bending cycles, showing less than 10% variation in responsivity after the first bending cycle of the as fabricated devices. **d**, Device responsivity as a function of annealing conditions of the array on both SiO2 and PI substrates. Annealing was done in an inert environment using a hotplate where arrays were allowed to heat up and cool down while on the plate. n = 16.

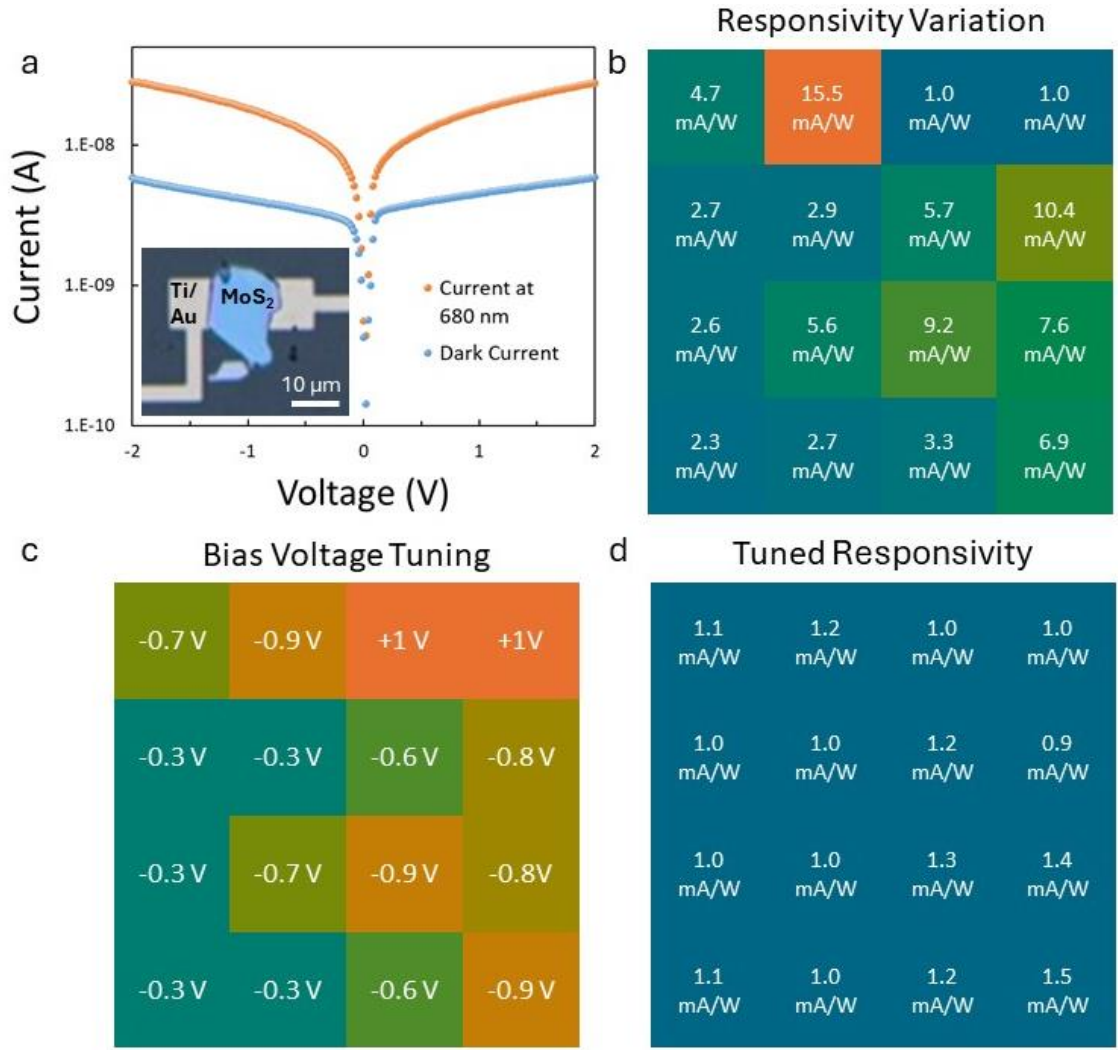

**Fig. 3:** Array variations and tuning. **a**, Example of device I-V characteristics with operation with dark current and light current operated at 8.2 µW and 680 nm. Absolute value of the device current is taken to plot on a log scale graph. Inset shows device 7 on PI, the device under test, with a 10 µm scale bar. **b**, As fabricated device characteristics for responsivities tested at 2 V, with colder colors representing lower device performance and warmer colors representing higher device performance. **c**, Tuning to 1 V bias voltage applied to each device to increase uniformity for the full array. Warmer colors represent larger absolute value for the device tuning, while colder colors represent lower absolute value for tuning. **d**, Array responsivities after applying voltage tuning from (c) minimal variation is seen between devices.

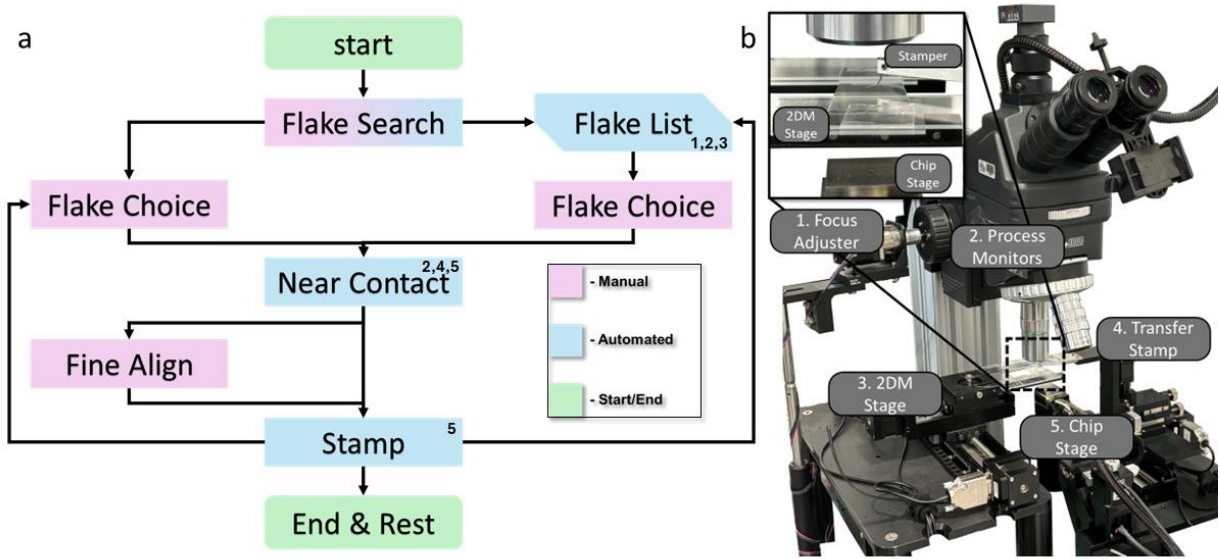

**Fig. 4:** 2D material transfer process and system. **a**, A flow chart for the 2D material transfer process showing automated and manual steps in the process where subroutines are marked with corresponding subsystems in (b). The process begins with a startup routine and then manually or automatically searching for suitable flakes that have been exfoliated on to a PDMS film, flakes must be manually chosen to be transferred to the given device region. Once the flake choice has been made the system will automatically do a rough alignment procedure and bring the flake to near contact with the device region, at which point the operator can choose to a manually fine alignment or to stamp the flake down for wider pitched devices. After the flake has been transferred the process can be repeated until all devices are created, at which point the system will run an automated end and rest routine. **b**, 2D material transfer system and labelled subsystems. The first subsystem is an automated focus adjuster which is used in conjunction with the process monitor subsystem consisting of a camera and image processing software designed to find when images of the PDMS sheet are in focus. Utilizing these two subsystems and the third subsystem, the 2DM stage, flakes can automatically be searched for and located using edge detection and centroid localization algorithm. The final two subsystems control the transfer stamper and chip stages, respectively, both subsystems have 3 dimensional movements with the vertical axis of the transfer stamper using a sensitive piezo-electric system.